\documentstyle[12pt]{article}
\begin{document}
\thispagestyle{empty}
\begin{center}
\LARGE \tt \bf {Faraday rotation and primordial magnetic fields constraints on ultraviolet Lorentz violation with spacetime torsion}
\end{center}

\vspace{3.5cm}

\begin{center}
{\large By L.C. Garcia de Andrade\footnote{Departamento de F\'{\i}sica Te\'{o}rica - IF - UERJ - Rua S\~{a}o Francisco Xavier 524, Rio de Janeiro, RJ, Maracan\~{a}, CEP:20550.e-mail:garcia@dft.if.uerj.br}}
\end{center}

\begin{abstract}
 Recently Kahniashivili et al (2006) presented a unified treatment for ultraviolet Lorentz violation (LV) testing through
electromagnetic wave propagation in magnetised plasmas, based on dispersion and rotation measured data. Based on the fact discovered 
recently by Kostelecky et al (2008), that LV may place constraints on spacetime torsion, in this paper it is shown that 
on the limit of very low frequency torsion waves, it is possible to constraint torsion from Faraday rotation and CMB on a similar fashion as 
Minkowski spacetime plus torsion. Here the Maxwells modified equations are obtained by a perturbative method introduced by de Sabbata
and Gasperini (1981). Torsion is constraint to $Q_{CMB}\approx{10^{-18}GeV}$ which is not so stringent as the $10^{-31}GeV$ obtained 
by Kostelecky et al. However, Gamma Ray Bursts (GBRs) may lead to the more string value obtined by Kostelecky et al.
Another interesting constraint on torsion is shown to be placed by galactic dynamo seed magnetic fields. For torsion effects 
be compatible with the galactic dynamo seeds one obtains a torsion constraint of $10^{-33}GeV$ which is two orders of 
magnitude more stringent that the above Kostelecky et al limit. 
\end{abstract}

\newpage

\section{Introduction}
\hspace{1.0cm}As recently pointed out by Gamboa et al \cite{1} the cornstone of cosmology is the cosmological principle which is based on the fact that 
the spacetime is Lorentz invariant. Nevertheless, physical problems such as that primordial magnetic fields (PMF) 
\cite{2}, matter antimatter asymmetry and dark energy (matter) lead us to rethink Lorentz invariant Einsteinian relativity dogma. One of the ways to address this issue
has been recently discovered by Kostelecky et al \cite{3}, which considers that LV could be associated to another alternative gravity theory
gravity theory called torsion theory \cite{4}. An important issue at this point is to stress that here as in most string inspired 
Kalb Rammond theory, torsion propagates in vacuum instead of being considered as a contact interaction as in Einstein-Cartan gravitation
\cite{5}. 
 In this paper we shall be concerned with some examples where LV not only is present in spacetime but it can be used to place
limits in spacetime torsion by its manifestations on dynamo effects and GBRs and cosmic microwave background (CMB)
with an interesting analogy of electromagnetic (EM) waves in magnetised plasmas \cite{6}. Faraday rotation, which is so
important in measuring magnetic fields \cite{7} is used to place constraints on LV torsion vector. An important issue is that here one addopts perturbative approach to quantum electrodynamics
(QED) effects and not the non-perturbative ones used by Enqvist et al \cite{8}. The idea of using this EW analogy in GRB 
ultraviolet LV has been used by Kahniatishvilly et al \cite{9} in Minkowski spacetime without torsion. One of the main diffrences 
between their results and ours, is that to EM waves frequency a torsion wave frequency is summed up, and only in the very low frequency 
torsion waves, they approach. In this first paper one takes in account this low frequency limit and left the high frequency torsion wave limit 
to a future work. One also consider here a linearised approach to dispersive and rotations measures. The paper is 
organised as follows: In section II the de Sabbata-Gasperini formulation of the Riemann-Cartan (RC) Maxwells 
vacuum electrodynamics, with photon-torsion semi-minimal coupling is reviewed. In section III the EM waves analogy in magnetised plasma is considered and 
the CMB torsion limit is placed in the low frequency torsion wave limit. In section IV galactic magnetic dynamo seeds are used to place limits on LV through
torsion modes. In section V conclusions and discussions are presented.
\section{Perturbative QED in Minkowski torsioned spacetime}
Throughout the paper second order effects on torsion 
shall be neglected in the electrodynamics including LV terms due to the three-dimensionl torsion vector $\textbf{Q}$. In this section we consider a 
simple cosmological application concerning the electrodynamics in vacuum QED 
spacetime background. Perturbative approach to electrodynamics leads to the following set of equations
\begin{equation}
{\partial}_{i}F^{ji}=4{\pi}J^{j}+\frac{2{\alpha}}{3{\pi}}{\epsilon}^{jklm}Q_{l}{F_{kj}}
\label{1}
\end{equation}
\begin{equation}
{\partial}_{[i}F_{jk]}=0
\label{2}
\end{equation}
where ${\partial}_{i}$ is the partial derivative, and $\alpha$ is the e.m fine structure constant, while $F^{ij}={\partial}^{i}A^{j}-{\partial}^{j}A^{i}$ 
is the electromagnetic field tensor non-minimally coupled to torsion gravity. Here $A^{i}$ is the electromagnetic 
vector potential and $(i,k=0,1,2,3)$ and $Q_{l}$ represents the torsion four-vector. In three-dimensional notation the above Maxwells generalised equations read
\begin{equation}
{\nabla}.{\textbf{E}}=4{\pi}{\rho}+\frac{4{\alpha}}{3{\pi}}\textbf{Q}.\textbf{B}
\label{3}
\end{equation}
\begin{equation}
{\nabla}.\textbf{B}=0
\label{4}
\end{equation}
\begin{equation}
{\nabla}{\times}\textbf{E}= -\frac{{\partial}{\textbf{B}}}{{\partial}t}
\label{5}
\end{equation}
\begin{equation}
{\nabla}\times{\textbf{B}}=\frac{4{\alpha}}{3{\pi}}\textbf{E}{\times}\textbf{Q}+\frac{{\partial}\textbf{E}}{{\partial}t}
\label{6}
\end{equation}
After some algebraic manipulation on these generalised Maxwell equations one obtains the EM wave equations
\begin{equation}
{\nabla}^{2}\textbf{E} -\frac{{\partial}^{2}{\textbf{E}}}{{\partial}t^{2}}+
\frac{4{\alpha}}{3{\pi}}[\textbf{Q}{\times}\frac{{\partial}\textbf{E}}{{\partial}t}-\textbf{E}{\times}\frac{{\partial}\textbf{Q}}{{\partial}t}]
=0\label{7}
\end{equation}
\begin{equation}
{\nabla}^{2}\textbf{B} -\frac{{\partial}^{2}{\textbf{B}}}{{\partial}t^{2}}-
\frac{16{\alpha}}{3}{\rho}\textbf{Q}-(\frac{4{\alpha}}{3{\pi}})^{2}\textbf{Q}(\textbf{Q}.\textbf{B})=0
\label{8}
\end{equation}
By Fourier analyzing the first expression or substituting ${\partial}_{t}\rightarrow{i{\omega}}$ and 
${\nabla}\rightarrow{-ik}$ one obtains from expression (\ref{7}) the following expression
\begin{equation}
[({\omega}^{2}-k^{2}){\delta}_{ab}-\frac{4{\alpha}}{3{\pi}}i({\omega}_{1}+{\omega}){\epsilon}_{acb}Q^{c}]E^{b}=0
\label{9}
\end{equation}
\begin{equation}
ik_{a}E_{a}=\frac{4{\alpha}}{3{\pi}}Q_{c}E_{c}
\label{10}
\end{equation}
where $(a,b=1,2,3)$, ${\omega}$ is the EM wave frequency while ${\omega}_{1}$ is the torsion wave frequency. Here we also chose the charge density 
${\rho}=0$ since we are addopting vacuum QED. The dispersion relation is given by
\begin{equation}
{\omega}^{2}\mp({\omega}_{1}+{\omega})kQ-k^{2}[1\mp{\gamma}]=0
\label{11}
\end{equation}
where ${\gamma}(k)$ is the photon-spin-sign-dependent term on the LHS of equation (\ref{11}), to account for the 
phenomenological LV of an energy-dependent photon speed. Now by considering the analogy to EM waves in a magnetised plasma with an index of refraction 
of refraction of $n=\frac{k}{\omega}$, one obtains
\begin{equation}
n_{L,R}=({\epsilon}_{1}\pm{\epsilon}_{2})^{\frac{1}{2}}
\label{12}
\end{equation}
where ${\epsilon}$ is the electric permittivity. From the dispersion relation above one obtains
the permittivities
\begin{equation}
{\epsilon}_{1}=\frac{1}{(1\pm{\gamma}(k))}
\label{13}
\end{equation}
\begin{equation}
{\epsilon}_{2}=-\frac{({\omega}_{1}+{\omega})Q}{(1\pm{\gamma}(k))}\approx{({\omega}+{\omega}_{1})}Q
\label{14}
\end{equation}
From the approximation of low torsion frequency ${\omega}_{1}<<<{\omega}$, this can be dropped in the last expression and torsion 
vector reduces to the $\textbf{g}$ LV vector used by Khniatishivilly et al. Thus these expressions one obtains the refraction index
\begin{equation}
{n}_{L,R}=(1\pm{\omega}Q\pm{\gamma}(k))^{\frac{1}{2}}
\label{15}
\end{equation}
By making the approximation ${\gamma}<<Q{\omega}$ the refractive index reduces to
\begin{equation}
{n}_{L,R}\approx{(1\pm{\omega}Q)^{\frac{1}{2}}}=\frac{k}{\omega}
\label{16}
\end{equation}
Now the dispersion measure and rotation measure (RM) of the GRBs depend on the photon travel distance ${\Delta}l$ and are expressed as
\begin{equation}
{\Delta}t_{L,R}={\Delta}l(1-\frac{{\partial}k_{L,R}}{{\partial}{\omega}})
\label{17}
\end{equation}
\begin{equation}
{\Delta}{\phi}=\frac{1}{2}(k_{L}-k_{R}){\Delta}l
\label{18}
\end{equation}
where ${\phi}$ is the polarization plane rotation of the electric field describing the Faraday rotation. 
These expressions can be written in terms of torsion by
\begin{equation}
{\Delta}t_{L,R}=\pm{\Delta}l{\omega}Q
\label{19}
\end{equation}
\begin{equation}
{\Delta}{\phi}\approx{\frac{1}{2}{\omega}^{2}Q{\Delta}l}
\label{20}
\end{equation}
Therefore, when the photon-spin is damped by the torsion wave, the Faraday rotation of 
${\Delta}{\phi}\approx{10^{-2}rad}$, allows one to estimate torsion as 
\begin{equation}
Q_{CMB}\approx{{10}^{-18}GeV}
\label{21}
\end{equation}
thus stablishing new limits for LV from torsin distinct from those of Kostelecky et al.
\section{Galactic dynamo seeds constraints to LV in spacetime with torsion}
Recently a more complicated approach to place constraints on LV from galactic dynamo magnetic seed fields appeared in 
the literature \cite{3,4}. Here folllowing the perturbative methd bove nd the magnetic field equation one obtains a much simpler 
and straightforward method of placing limits on LV from torsion and galactic dynamo seeds. Performing the Fourier spectrum of the
magnetic field equation yields
\begin{equation}
[({\omega}^{2}-k^{2}){\delta}_{ab}+\frac{16{\alpha}^{2}}{9{\pi}}Q_{a}Q_{b}]B^{b}=0
\label{22}
\end{equation}
which yields the following dispersion relation
\begin{equation}
{\omega}^{2}= k^{2}+\frac{16{\alpha}^{2}}{9{\pi}^{2}}Q^{2}
\label{23}
\end{equation}
Actually ${\omega}$ coincides with dynamo growth rate ${\gamma}_{0}$ with the ansatz
\begin{equation}
B(t)= B_{0}e^{{\gamma}t}
\label{24}
\end{equation}
From the dispersion relation one may conclude that in order that torsion may contribute to dynamo action it must be 
be comparable with the large scale coherence which is the inverse of the wave vector k, under the law
\begin{equation}
k^{2}\approx{\frac{16{\alpha}^{2}}{9{\pi}^{2}}Q^{2}}
\label{25}
\end{equation}
As for today coherence scales torsion would be extremely weak and of the order of $Q\approx{10^{-21}cm^{-1}}$. 
This is exactly the estimate obtained by Laemmerzahl \cite{11} on the basis of Earth laboratory Hughes-Drever experiment. It is interesting to note that Kostelecky et al 
have also obtained LV with table top experiments on Earth lab.
\section{Discussion and conclusions}
The investigation of Faraday rotation has been proved very important in high-energy astronomy of magnetic fields in the 
universe. Here one uses Faraday rotation to stablish limits of the LV in terms of torsion as stblished by Kostelecky 
et al with torsion being a constant vector. Actually here torsion vector is not constant though LV is attainble. Methods used here were previously 
investigated by Kahniatshivilly et al in the context of GRBs in torsionless Minkowski spacetime. 
Dynamo plasma is obtained from the dispersion relation where torsion can be express3ed in terms of the coherence 
scale of magnetic fields. Quantum effects may be obtained here from perturbative QED instead of non-perturbative primordial magnetic fields obtained by Enqvist.  
\section*{Acknowledgements} 
 I would like to express my gratitude to Tina Kahniashivily and Rodion Stepanov for helpful discussions on the 
subject of this paper. I also thank Tanmay Vachaspati for his kind invitation to the Primordial magnetic fields 2011 held in rizona State 
University last april. Thanks are due to F Hehl for some useful  correspondence on torsion waves. I also benifit from discussions with some participants like Bharat Ratra. Financial support from CNPq. and University of State of Rio de Janeiro (UERJ) are grateful acknowledged.

\end{document}